\documentclass[aps,prb,preprint,superscriptaddress]{revtex4-2}
\usepackage{graphicx}  
\usepackage{dcolumn}   
\usepackage{bm}        
\usepackage{amssymb}
\usepackage[english]{babel}
\usepackage{booktabs}
\usepackage[export]{adjustbox}
\usepackage{array}
\usepackage{subcaption}
\usepackage{mathtools}
\usepackage{amsmath}
\usepackage{xcolor}

\begin{document}
\title{Emission enhancement and bandgap narrowing in $Cs_2TeBr_6$ under pressure}

\author{Debabrata Samanta}
\email [Present Address:]{ Beijing Academy of Quantum Information Sciences, Beijing, 100193, P.R. China}
\affiliation{National Centre for High Pressure Studies, Department of Physical Sciences, Indian Institute of Science Education and Research Kolkata, Mohanpur Campus, Mohanpur 741246, Nadia, West Bengal, India.}

\author{Suvashree Mukherjee}
\affiliation{National Centre for High Pressure Studies, Department of Physical Sciences, Indian Institute of Science Education and Research Kolkata, Mohanpur Campus, Mohanpur 741246, Nadia, West Bengal, India.}

\author{Asish Kumar Mishra}
\affiliation{National Centre for High Pressure Studies, Department of Physical Sciences, Indian Institute of Science Education and Research Kolkata, Mohanpur Campus, Mohanpur 741246, Nadia, West Bengal, India.}

\author{Bhagyashri Giri}
\affiliation{National Centre for High Pressure Studies, Department of Physical Sciences, Indian Institute of Science Education and Research Kolkata, Mohanpur Campus, Mohanpur 741246, Nadia, West Bengal, India.}

\author{Sonu Pratap Chaudhary}
\email [Present Address:]{ Department of Applied Science \& Humanities, Goverment Engineering College Buxar (Science, Technology and Technical Education Department, Goverment of Bihar), Buxar 802103, India.}
\affiliation{Department of Chemical Sciences, and Centre for Advanced Functional Materials, Indian Institute of Science Education and Research (IISER) Kolkata, Mohanpur-741246, India.}

\author{Konstantin Glazyrin}
\affiliation{Photon Science, Deutsches Elektronen Synchrotron, Hamburg, Germany}

\author{Sayan Bhattacharyya}
\affiliation{Department of Chemical Sciences, and Centre for Advanced Functional Materials, Indian Institute of Science Education and Research (IISER) Kolkata, Mohanpur-741246, India.}

\author{Goutam Dev Mukherjee}
\email [Corresponding author:]{goutamdev@iiserkol.ac.in}
\affiliation{National Centre for High Pressure Studies, Department of Physical Sciences, Indian Institute of Science Education and Research Kolkata, Mohanpur Campus, Mohanpur 741246, Nadia, West Bengal, India.}
\date{\today}

\begin{abstract} 
Pressure-induced emission enhancement and bandgap narrowing in vacancy-ordered halide double perovskite $Cs_2TeBr_6$ are extensively investigated through photoluminescence and absorption experiments. The below bandgap broad emission is attributed to self-trapped excitons recombination.  The $Cs_2TeBr_6$ crystal, consisting of undistorted octahedra, exhibits substantial emission enhancement due to the lowering of the energy barrier between $^3P_1$ and  self-trapped exciton states, as well as the suppression of nonradiative energy loss with increasing pressure.  In the Raman measurements, the observed behavior of full width at half maximum of all Raman modes implies dominant electron-phonon interactions rather than anharmonic interactions between phonons. The pressure-dependent X-ray diffraction measurements reveal an anomalous behaviour in the normalized pressure as a function of the Eulerian strain.

\end{abstract}

\maketitle


In recent years, there has been a lot of interest in researching semiconducting materials for optoelectronic applications. In this search, vacancy-ordered halide double perovskites (VOHDPs) of the family $Cs_2MX_6$ (M=Sn, Te, Ti, Zr; X= Cl, Br, I) have generated enormous interest due to their intrinsic optoelectronic properties \cite{Cai}. The importance of these materials demands extensive investigation under external perturbations (pressure, temperature, etc.) before their application in optoelectronic devices. The strain induced by hydrostatic pressure has been found to modify the structural, optical, and electronic properties of halide perovskites \cite{Samanta, Samanta1}. Very few works have focused on the structural, optical, and electronic properties of VOHDPs under pressure \cite{Bounos, Liu}. For examples, $Cs_2SnCl_6$ and  $Cs_2SnBr_6$ crystals retain cubic structure upon compression up to 20 GPa \cite{ Bounos}. Under pressure, the absorption coefficient of $Cs_2TiBr_6$ increases, and the absorption edge shifts towards lower energy \cite{Liu}. G. Bounos et al. \cite{ Bounos} observed dramatic changes in Raman spectra for $Cs_2SnI_6$ at 8 GPa, corresponding to a cubic to monoclinic structural phase transition confirmed by synchrotron X-ray diffraction (XRD). In  $Cs_2TeCl_6$, self-trapped excitons (STEs) emission enhances with increasing pressure up to 1.6 GPa and two electronic transitions are observed at 1.6 and 5.8 GPa, respectively \cite{Shi}.

The  $Cs_2TeBr_6$ crystal, a member of the aforenamed family, crystalizes in cubic structure (Fm$\bar{3}$m) \cite{Das} and is a suitable material for next-generation sensory devices \cite{Li}. $Cs_2TeBr_6$ single crystal exhibits a broad emission corresponding to the $^3P_1$ $\rightarrow$ $^1S_0$ transition of $Te^{4+}$ ion \cite{Folgueras}. $Cs_2TeBr_6$ microcrystals have demonstrated high photocatalytic activity \cite{Huang}. B. Cucco et al. \cite{Cucco}	have revealed how electron-hole interactions affect optical properties and predicted small effective masses ($m_e$=0.33$m_0$ and $m_h$=0.46$m_0$) in $Cs_2TeBr_6$. A bulk modulus of 15.83 GPa and an indirect bandgap of 1.88 eV have also been predicted by a first-principle study \cite{Sa}. However, as far as our knowledge experimental investigations are lacking 
   
Under pressure, VOHDPs undergo structural phase transitions with distinct electronic and transport properties. Strain can alter the scattering mechanisms: phonon-phonon and electron-phonon interactions. Both can influence a phonons' lifetime.  The latter affects the mobility of charge carriers and, consequently, the materials' conductivity. The band gap change due to chemical strain modifies the absorption and emission spectra \cite{Folgueras}. Similar to chemical strain, mechanical strain can also be utilized to engineer the material's optical properties for specific applications. Therefore, understanding the correlation among structural, optical, and electronic properties of  $Cs_2TeBr_6$ under  mechanical strain is of great importance for future applications in optoelectronics.
 
 In this work,  pressure-induced emission enhancement and bandgap narrowing in $Cs_2TeBr_6$ are extensively investigated through photoluminescence and absorption experiments. The $Cs_2TeBr_6$ crystal, consisting of undistorted octahedra, exhibits substantial emission enhancement due to the lowering of the energy barrier between $^3P_1$ and  self-trapped exciton states, as well as the suppression of nonradiative energy loss with increasing pressure. \\

To synthesize $Cs_2TeBr_6$ powder, a 2:1 molar ratio of $CsBr$ and $TeO_2$ is dissolved in 2 mL $HBr$ acid in a vial. The mixture is heated to 120$^o$ C and stirred to react for 2 hours. After that, the mixture is kept for an additional 2 hours at the same temperature without stirring. The obtained solution is allowed to cool to room temperature naturally overnight. The precipitated $Cs_2TeBr_6$ compound is washed with ethanol and dried for experimental use.
The characterization  of the synthesized $Cs_2TeBr_6$ powder is performed by
synchrotron XRD measurements. The ambient sample XRD pattern is indexed to a cubic structure ( space group: Fm$\bar{3}$m) with lattice parameter $a=10.8690(6)  \AA$. Rietveld refinement plot of the ambient XRD pattern is enclosed in Fig. 1(a). Atomic positions of $Cs_2TeBr_6$ include Te on 4a (0, 0, 0), Cs on 8c (0.25, 0.25, 0.25), and Br on 24e [0.2490(4), 0, 0], which are in good agreement with previous literature \cite{Das}. Fig. 1(b) shows the crystal structure of  $Cs_2TeBr_6$.  The unit cell consists of isolated undistorted octahedra. Te atoms are located at the center of octahedra. Br atoms occupy octahedra corners and octahedra are separated by Cs atoms.

We have employed a piston-cylinder type diamond anvil cell (DAC) to generate pressure for Raman scattering, photoluminescence, and absorption experiments. High-pressure PL measurements up to 9.2 GPa are carried out using a Monovista micro-Raman spectrometer equipped with a 750 mm monochromator and a back-illuminated PIXIS 256 CCD detector with thermoelectric cooling down to -70$^o$C. A laser of wavelength 532 nm is used as an excitation source. The pressure evolution of PL spectra in Figures 3(a) and 3(b) shows a drastic change in intensity and peak position.  At ambient conditions, a broad PL spectrum centered at 696 nm and spreading from visible to near-infrared is observed. The ambient PL profile shown in Fig. S1 is centered at 696 nm with a full width at half maximum (FWHM) of 137 nm and a long tail on the long wavelength side. In a previous report \cite{Folgueras}, the PL has been attributed to STEs emission. The relaxation occurs from the excited states of Te$^{4+}$ through  $^3P_1$ $\rightarrow$ $^1S_0$ transition via STEs state. To investigate emission behaviour under pressure, PL peak position, integrated intensity, and FWHM are estimated by fitting each spectrum to a Voigt function, the convolution of a Gaussian and a Lorentzian function. With increasing pressure, the PL peak position exhibits a blue shift, shown in Fig. 3(c).  PL intensity generally weakens on compression. As presented in Fig. 1(d), PL intensity shows a maximum at around 2.7 GPa followed by a decrease until it ceases. The relative PL integrated intensity increases 20 times by applying 2.7 GPa pressure only. The pressure variation of FWHM of PL is displayed in the inset of Fig. 3(d). The narrowing of PL indicates a decrease in electron-phonon coupling with increasing pressure.

High-pressure absorption measurement are performed using the Sciencetech 15189RD custom made absorption spectrometer. In Fig. 4(a), the ambient PL and absorbance spectra exhibit a 0.7 eV Stokes shift, the characteristics of the STE emission. The absorbance is calculated as: A$=-\log(\frac{I_t-I_d}{I_0-I_d})$, where $I_0$ is the intensity of input light, $I_t$ is the intensity of light transmitted through the sample, and $I_d$ is the intensity at dark environment. The bandgap is determined from the absorption edge using a Touc plot for indirect bandgap [Fig. S2].  The linear absorption edge is fitted by the equation: $ (A h \nu)^{0.5} =C \times (h \nu -E_g)$;  where $A$ is the absorbance, $h \nu$ is the energy of the photon, $\it C$ is a constant, and $E_g$ is the indirect optical bandgap. The bandgap is measured to be  $E_g$=1.9 eV, which is closer to the previously reported bandgap values predicted by density functional theory \cite{Sa}.  Absorption spectra at various pressures are shown in Fig. 4(b). As pressure increases, the absorption edge shifts towards lower energy.   Figure 4(c) depicts the variation in bandgap with pressure. Upon compression from ambient to 9.5 GPa, the bandgap reduces by 10\%, while almost in the same pressure range the PL peak shifts by 6\% only. 
 
 Pressure-dependent Raman spectra are recorded using a Monovista micro-Raman spectrometer with an excitation source of wavelength 532 nm. To collect Raman spectra from 10 $cm^{-1}$, we have used a Bragg filter with a grating of 1500 $gr/mm$. A laser source of wavelength 532 nm is used as an excitation for Raman scattering measurement. The  $Cs_2TeBr_6$ crystal shows four Raman active modes $P_1 (T_{2g-Cs})$, $P_2 (T_{2g})$, $P_3 (E_g)$, and $P_4 (A_{1g})$ at 41.8, 87.5, 145.6, and 169.2 $cm^{-1}$, respectively [Fig 4(a)]. $P_3$ and $P_4$ Raman modes originate due to asymmetric and symmetric stretching vibration of $Te-Br$ bonds, respectively while bending vibration of $TeBr_6$ corresponds to  $P_2$ mode \cite{Stefanovich}. The $P_1$ mode is assigned to the translational motion of $Cs^{+}$ ion \cite{Folgueras}.  Raman spectra at selected pressures are shown in Fig. 4(a). Intensity of all Raman modes become very low at 10.2 GPa and eventually merge with the background at 12.8 GPa. As shown in Fig. 4(b), upon hydrostatic compression, all Raman modes exhibit blue shift due to the decrease in interatomic distances accompanied by a decrease in unit cell volume. The slope of $P_1$  mode is 2.9 $cm^{-1}/GPa$ below 2.7 GPa and decreases to 1.6 $cm^{-1}/GPa$ above 2.7 GPa. This indicates that  the sample is highly compressible in the lower pressure region. Surprisingly, the intensity of the $P_1$ mode increases with an increase in pressure [Fig. 4(c)]. 
 The pressure dependence of FWHM of all Raman modes is shown in Fig. 4(d). $P_2$ mode becomes narrower,  $P_1$ and $P_3$ modes remain almost constant, $P_4$ mode narrowers until 2.2 GPa then broadens with increasing pressure.
 
 In Fig. 5(a), pressure-dependent X-ray diffraction (XRD) patterns show no changes up to 10.3 GPa. Thus, no phase transition is observed up to the highest pressure of this experiment. High pressure continuously contracts lattice parameters, as evidenced by Fig. 5(b). Volume vs. pressure data are fitted to the 3$^{rd}$ order Birch-Murnaghan equation of state \cite{Birch, Ahmad} to measure bulk modulus (B$_0$) and its pressure derivative (B$^{'}$). The fit results B$_0$=14$\pm$2 GPa and B$^{'}$=6.6$\pm$0.7 [Fig.5(c)]. The low bulk modulus indicates a soft lattice. To investigate the effect of internal strain on PL, we have plotted normalised pressure ($H$) vs. Eulerian strain ($f_E$) in Fig. 5(d), showing a minimum at around 2.7 GPa in the absence of structural phase transition. The higher pressure region (above 2.7 GPa) is fitted to the equation, $H=B_0+\frac{3}{2}B_0(B^{'}-4)f_E$ \cite{Polian}. The fitted parameters: $B_0$=15.57$\pm$0.52 GPa and $B^{'}$=6.45$\pm$0.38. In the lower pressure region (below 2.7 GPa), $B^{'}$ is less than 4 and the sample is highly compressible.  The pressure variation of compressibility is shown in Fig. S5.

Hydrostatic pressure influences electron-phonon interactions, effect of which  is evident in pressure-dependent Raman and  PL measurements. 
The Raman mode frequency shift primarily results from anharmonic interactions between phonons and electron-phonon coupling. In the lower pressure region, the larger frequency shift for $P_1$  mode indicates higher compressibility, which is further confirmed by the XRD measurements. The strong anharmonic interaction typically leads to softening of a Raman mode under pressure \cite{Pawbake}, which is not observed in our experiment.  The observed behavior of FWHM of all Raman mode, shown in Fig. 4(d), implies dominant electron-phonon interactions rather than anharmonic interactions between phonons. 
The soft lattice with adequate electron-phonon coupling plays a crucial role in forming the STE state. During photo-excitation, excited carriers deform the lattice and form STE states inside the bandgap. As the STE states are the midgap states, they are less affected by external pressure. Therefore, the PL peak shift is smaller than the bandgap shift in the same pressure range [Fig. 3(c)]. The emission enhancement in halide perovskites has mostly been explained by octahedral distortion under high pressure \cite{Zou}. Surprisingly, the $Cs_2TeBr_6$ crystal, consisting of undistorted octahedra, exhibits substantial emission enhancement at higher pressures.  The enhanced emission can be attributed to the lowering of the energy barrier between $^3P_1$ and STE state, as well as the suppression of nonradiative energy loss with increasing pressure. As illustrated in Fig. 3(c), at an intermediate pressure around 2.7 GPa, the PL peak position coincides with the bandgap, hence the PL intensity reaches its maximum.  The FWHM of PL, which measures the strength of electron-phonon coupling, decreases with compression. When electron-phonon coupling is reduced, the depth of the STE state decreases, causing the blue shift of the PL peak. The exciton binding energy reduces up to 2.7 GPa, evidenced by Fig. 3(c). As the lattice is compressed under pressure, the electronic band structure changes, which in turn modifies the effective mass of excitons and hence the exciton binding energy.

In conclusion, the below bandgap broad emission of  $Cs_2TeBr_6$  is attributed to self-trapped excitons recombination.  The pressure-induced emission enhancement is ascribed to the lowering of the energy barrier between $^3P_1$ and  self-trapped exciton states, as well as the suppression of nonradiative energy loss with increasing pressure.  This work will accelerate further research  to understand pressure-induced emission enhancement on VOHDPs, consisting of undistorted octahedra. Our results also show a way to improve optical properties for optoelectronic applications. We believe future research will make effort to compensate the hydrostatic pressure-induced strain in the structure by alloying.

See the supplementary material containing information on experimental procedures, PL, Raman, absorption, and XRD data analysis, as well as other relevant information.

DS acknowledges the fellowship grant provided by the INSPIRE program, Department of Science and Technology, Government of India. GDM acknowledges the financial support from the Science and Engineering Research Board (SERB), Government of India, Grant No. CRG/2021/004343 for the development of the lab based high-pressure UV-VIS-NIR Absorption Spectrometer. The financial support from the Ministry of Earth Sciences, Government of India, grant number MoES/16/25/10-RDEAS is gratefully acknowledged. We also acknowledge the financial support for the India@DESY collaboration under the Department of Science and Technology, Government of India to carry out high-pressure XRD measurement at the P02.2 beamline of PETRA III, DESY, Germany.


\begin{figure}[h]
	\centering
	\includegraphics[scale = 0.7]{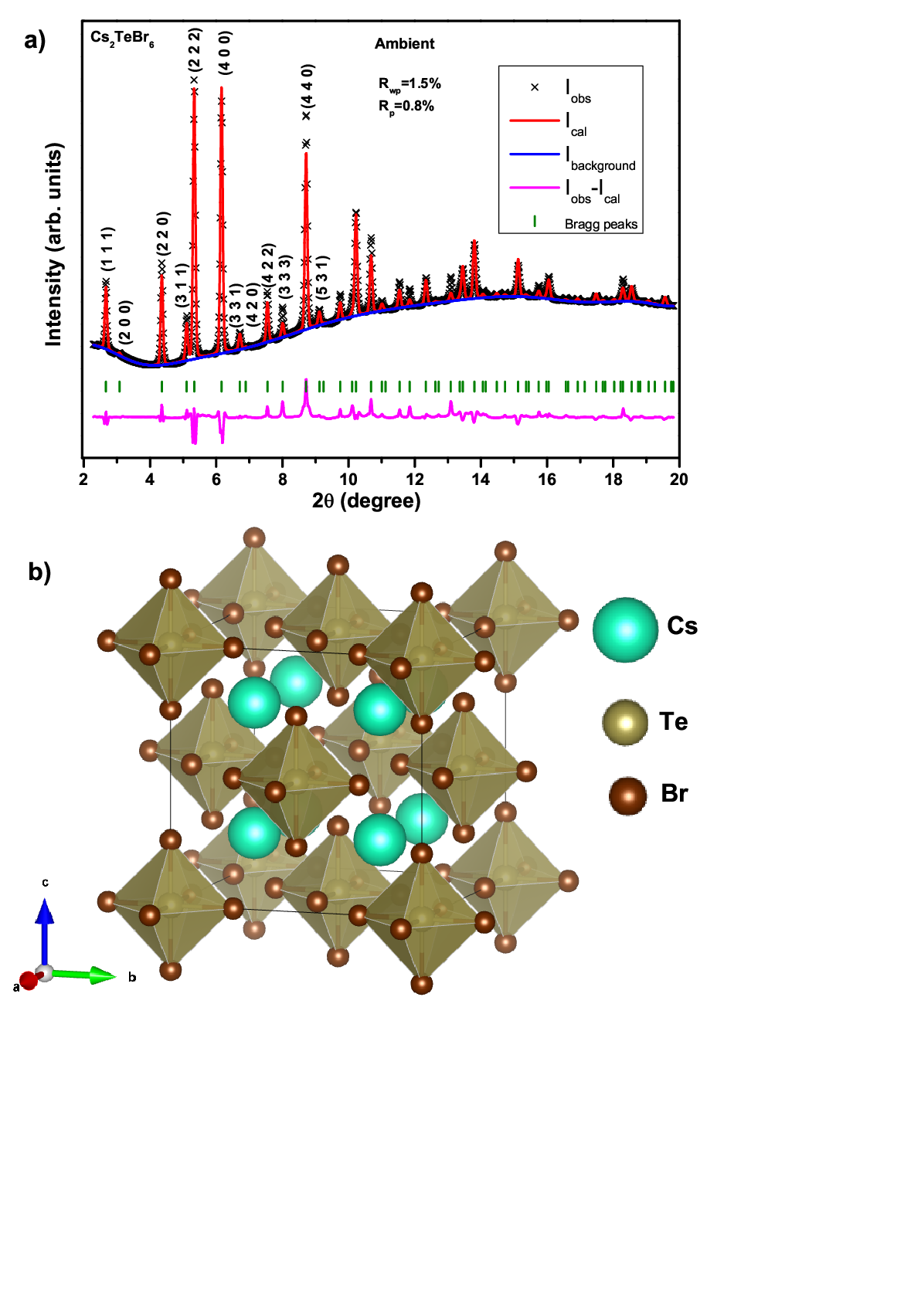}
	\vspace*{-35mm}
	\caption*{FIG. 1.The characterization  of the synthesized $Cs_2TeBr_6$ powder is performed by synchrotron XRD measurement.  (a) Rietveld plot of the ambient XRD pattern. The  $Cs_2TeBr_6$ crystal holds Fm$\bar{3}$m symmetry with lattice parameter $a=10.8690(6)  \AA$. Atomic positions of $Cs_2TeBr_6$ include Te on 4a (0, 0, 0), Cs on 8c (0.25, 0.25, 0.25), and Br on 24e [0.2490(4), 0, 0]. (b) The visualization of the unit cell of $Cs_2TeBr_6$.}
\end{figure}


\begin{figure}[h]
	\centering
	\includegraphics[scale = 0.7]{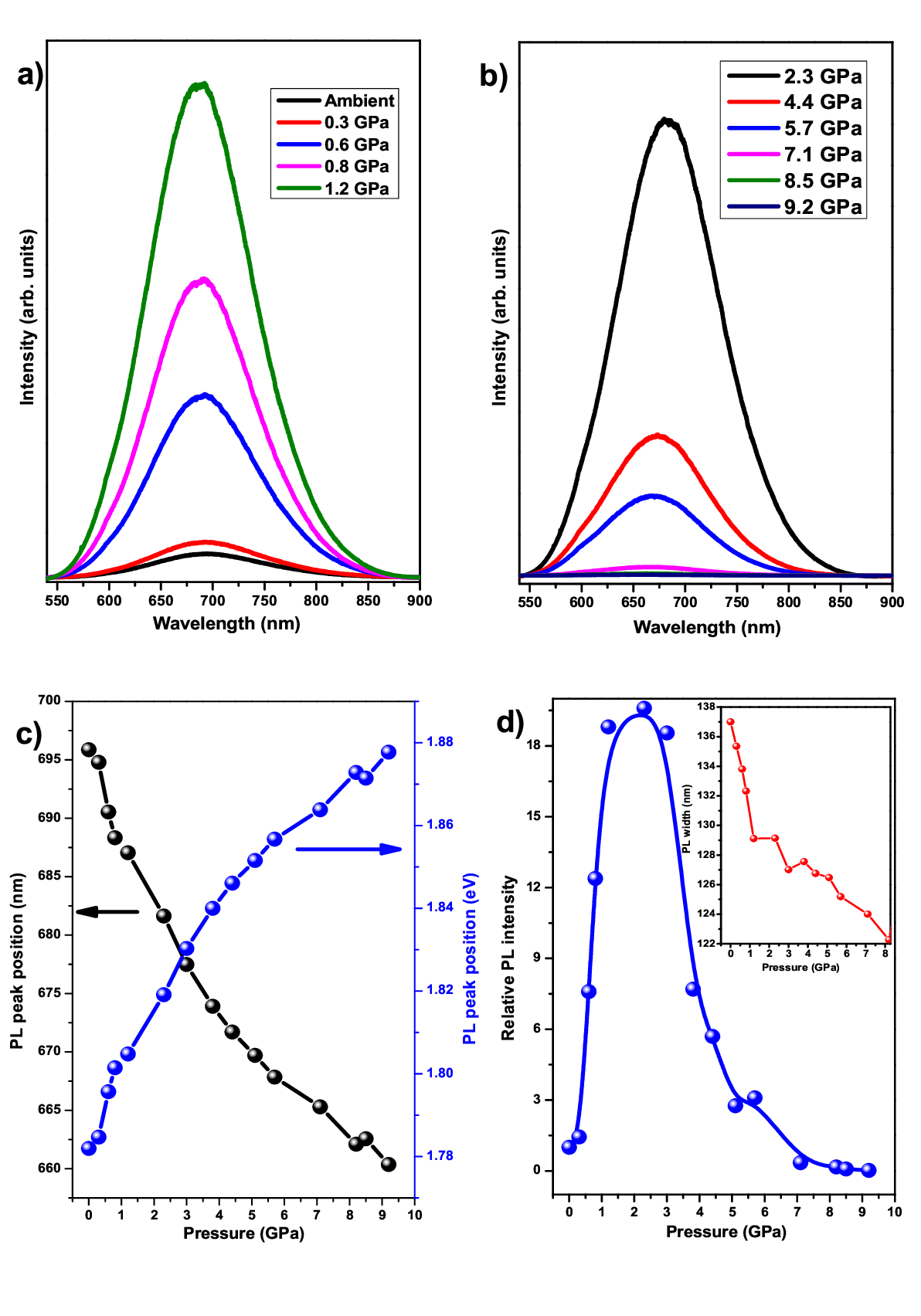}
	\caption*{FIG. 2. PL spectra of $Cs_2TeBr_6$ are recorded using an excitation of wavelenght 532 nm. (a) and (b) PL spectra at selected pressure poits. The evolution of (c) PL peak position and (d) relative PL intensity with pressure.}
\end{figure}
\begin{figure}[h]
	\centering
	\includegraphics[scale = 0.7]{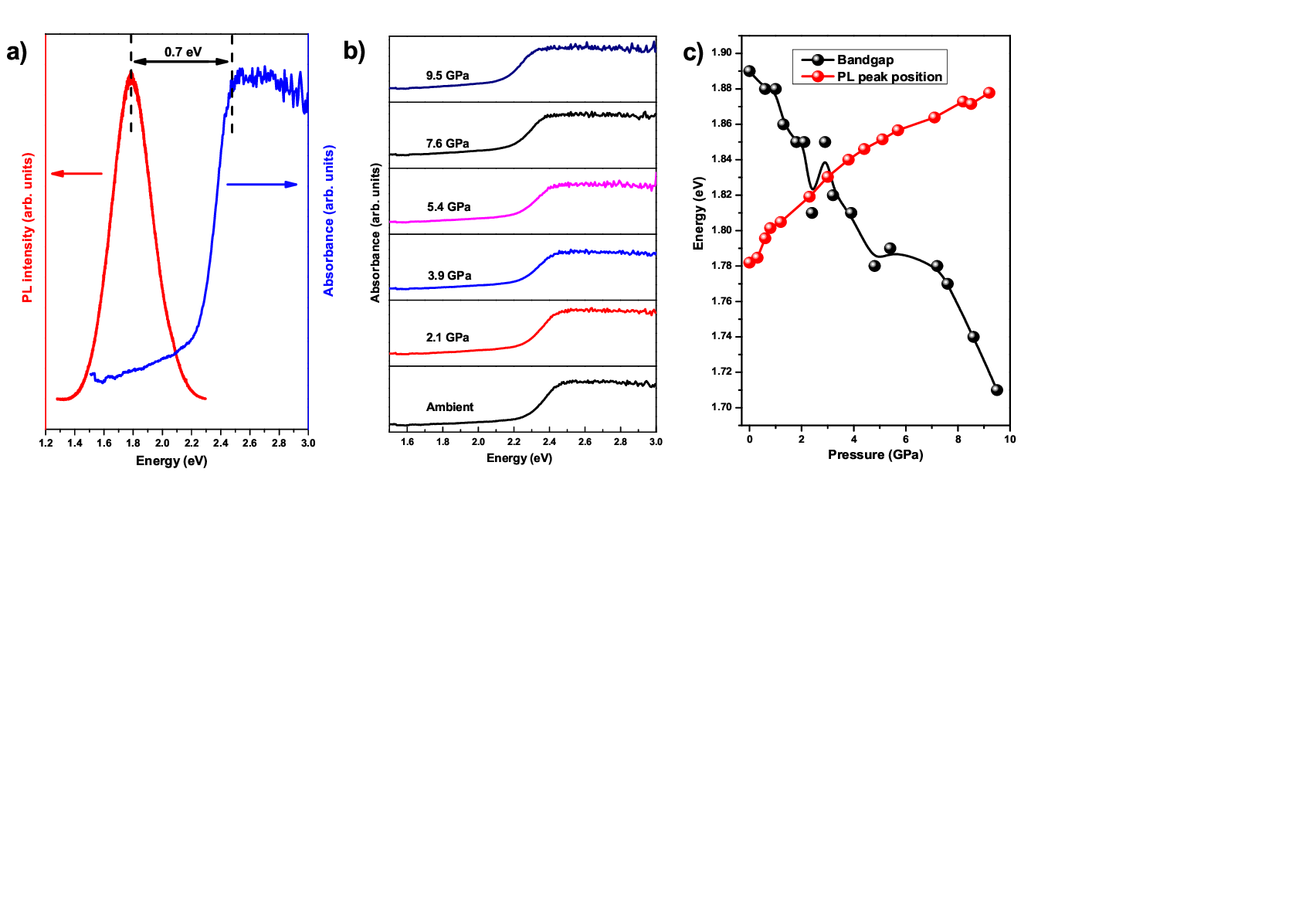}
	\vspace*{-70mm}
	\caption*{FIG. 3. (a) PL and absorption spectra of $Cs_2TeBr_6$. (b) UV-Vis absorption spectra at selected pressures. (c) The evolution of the indirect bandgap  and PL peak position as a function of pressure.}
\end{figure}

\begin{figure}[h]
	\centering
	\includegraphics[scale = 0.55]{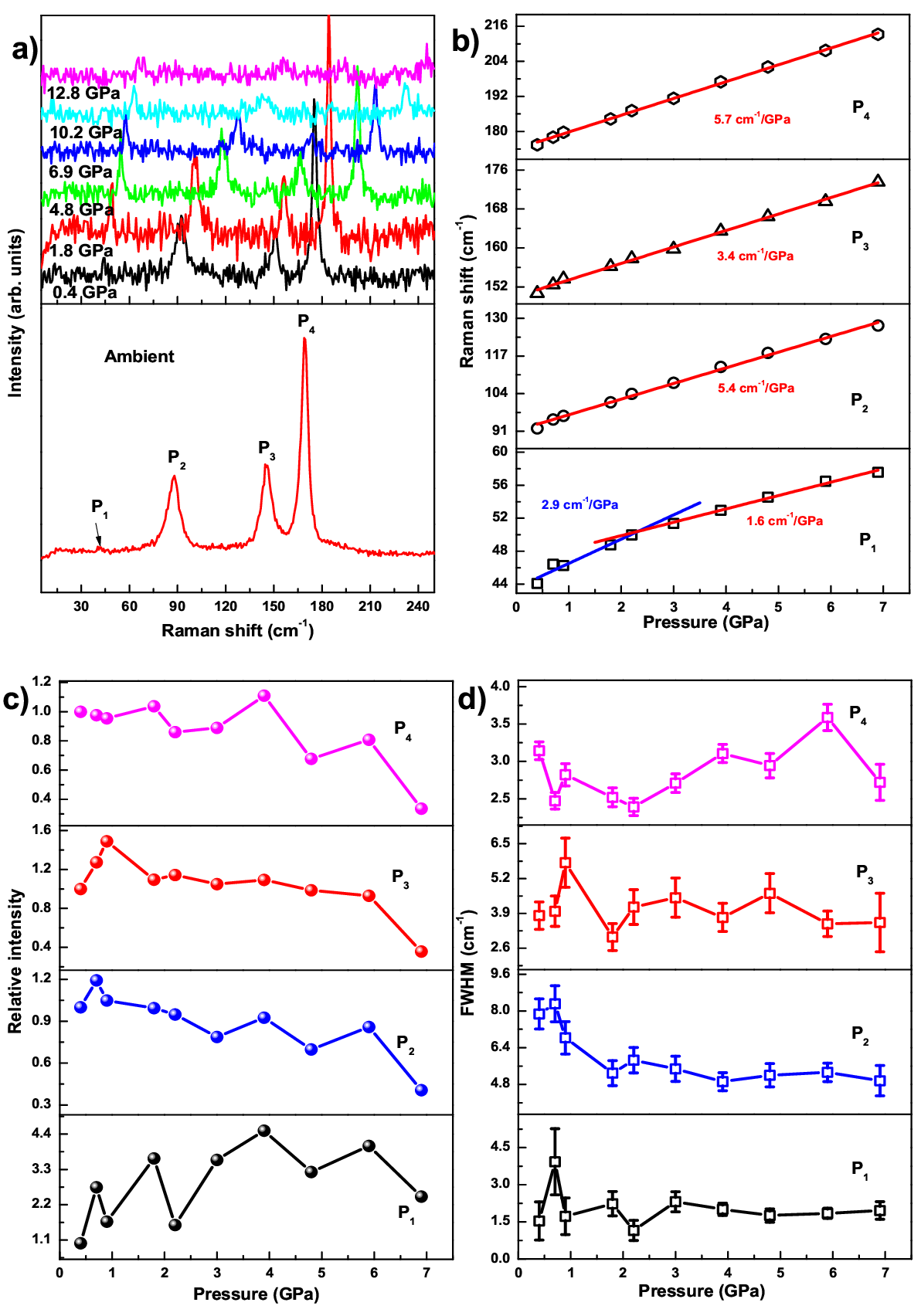}
	\caption*{FIG. 4. Pressure-dependent Raman scattering measurement of $Cs_2TeBr_6$. (a) Raman spectra at selected pressure points. The variation of (b) friequencies (c) relative intensity, and (d) FWHM of Raman modes as a function of pressure.}
\end{figure}

\begin{figure}[h]
	\centering
	\includegraphics[scale = 0.7]{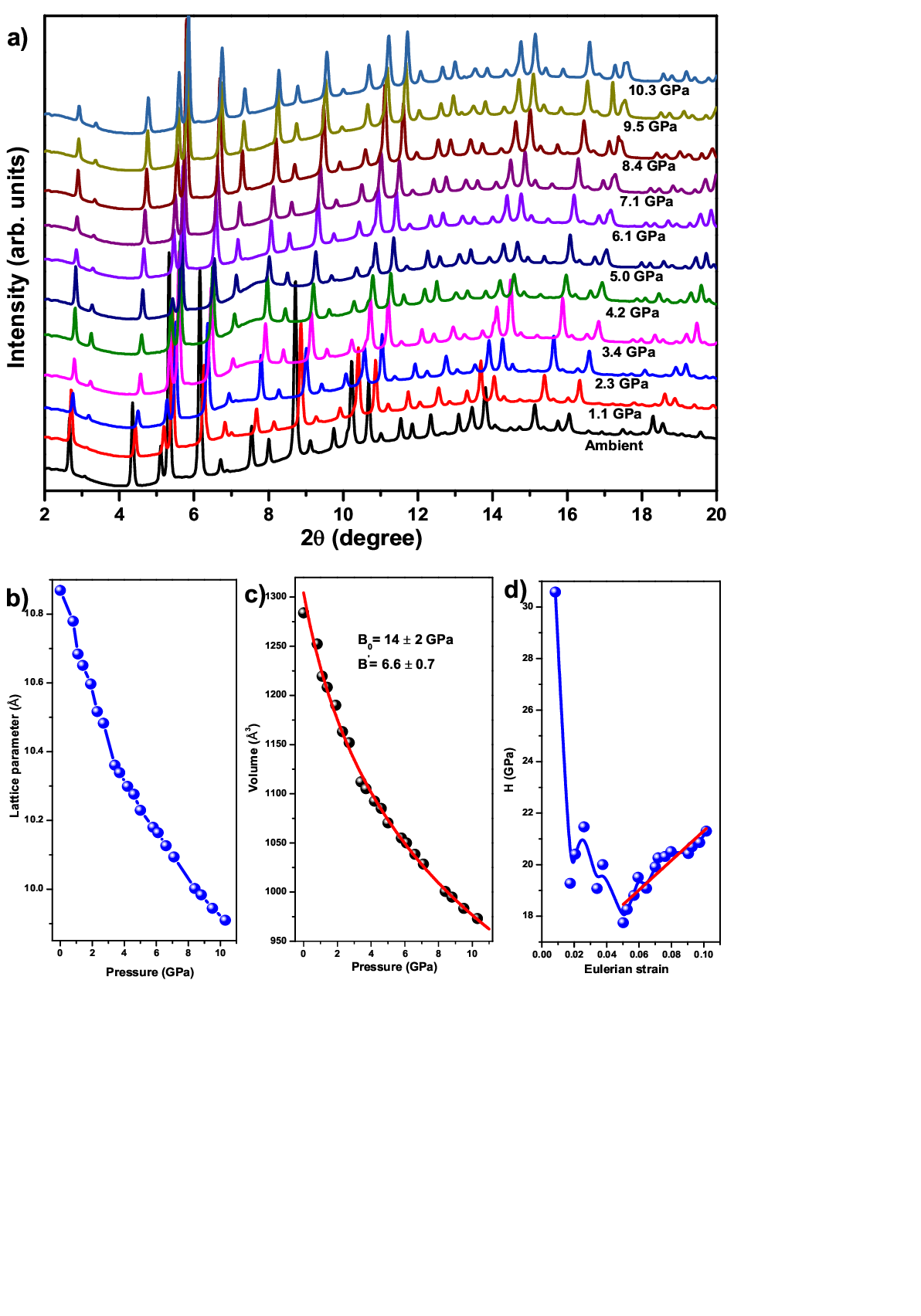}
	\vspace*{-35mm}
	\caption*{FIG. 5. High-pressure XRD measurements are carried out using monochromatic synchrotron X-ray of wavelength 0.29 $\AA$. (a)  XRD patterns of $Cs_2TeBr_6$. The evolution of (b) lattice parameters and (c) unit cell volume with varying pressure. (d) A plot of normalised pressure ($H$) vs. Eulerian strain ($f_E$).}
\end{figure}

\newpage
\noindent Additional figures and tables in support of the analyses carried out are presented here.
\section*{I. EXPERIMENTAL SECTION}
To synthesize $Cs_2TeBr_6$ powder, a 2:1 molar ratio of $CsBr$ and $TeO_2$ is dissolved in 2 mL $HBr$ acid in a vial. The vial is kept in a silicon oil bath to maintain homogeneous heating.  A thermometer is immersed in the silicon oil bath for accurate temperature measurement. The mixture is heated to 120$^o$ C and stirred to react for 2 hours. After that, the mixture is kept for an additional 2 hours at the same temperature without stirring. The obtained solution is allowed to cool to room temperature naturally overnight. The precipitated $Cs_2TeBr_6$ compound is washed with ethanol and dried for experimental use.

We have employed a piston-cylinder type diamond anvil cell (DAC) to generate pressure for Raman scattering, photoluminescence, and absorption experiments. A 290 $\mu m$ thick steel gasket  is preindented to 50 $\mu m$ by compressing it between two diamonds. A 100 $\mu m$ hole is drilled in the center of the indented portion. The sample, pressure marker, and pressure transmitting medium (PTM) are then loaded into the sample chamber with a diameter of 100 $\mu m$. Pressure is measured using the Ruby fluorescence technique \cite{Mao}, and silicon oil is used as a PTM. Raman and PL spectra are recorded using a Monovista micro-Raman spectrometer equipped with a 750 mm monochromator and a back-illuminated PIXIS 256 CCD detector with thermoelectric cooling down to -70$^o$C. A laser source of wavelength 532 nm is used as an excitation for both Raman and PL measurements. An infinity-corrected 20X objective focuses the incident radiation beam to sample and collects scattered radiations. To collect Raman spectra from 10 $cm^{-1}$, we have used a Bragg filter with a grating of 1500 $gr/mm$. High-pressure absorption measurement are performed using the Sciencetech 15189RD custom made absorption spectrometer. To focus broadband light on the sample, we have used an achromatic lens with a focal length of 19 mm. The transmitted light is collected through a 10X objective and recorded  using Silicon detector.
 
 \textcolor{blue} {High-pressure synchrotron X-ray diffraction (XRD) measurements are carried out at the P02.2 beamline of PETRA III, DESY, Germany with monochromatic X-ray of wavelength 0.2907 $\AA$. The pressure is varied using a membrane-driven symmetric DAC, with Neon as the PTM. Ruby fluorescence technique \cite{Mao} is employed to calibrate the pressure. Acquired two-dimensional diffraction images are converted to intensity versus 2$\theta$ profile using Dioptas software \cite{Preschera}. All XRD data are analyzed using the Rietveld refinement program implemented in the GSAS software \cite{Toby}.}
 
\begin{figure}
	\centering
	\includegraphics[scale = 0.6]{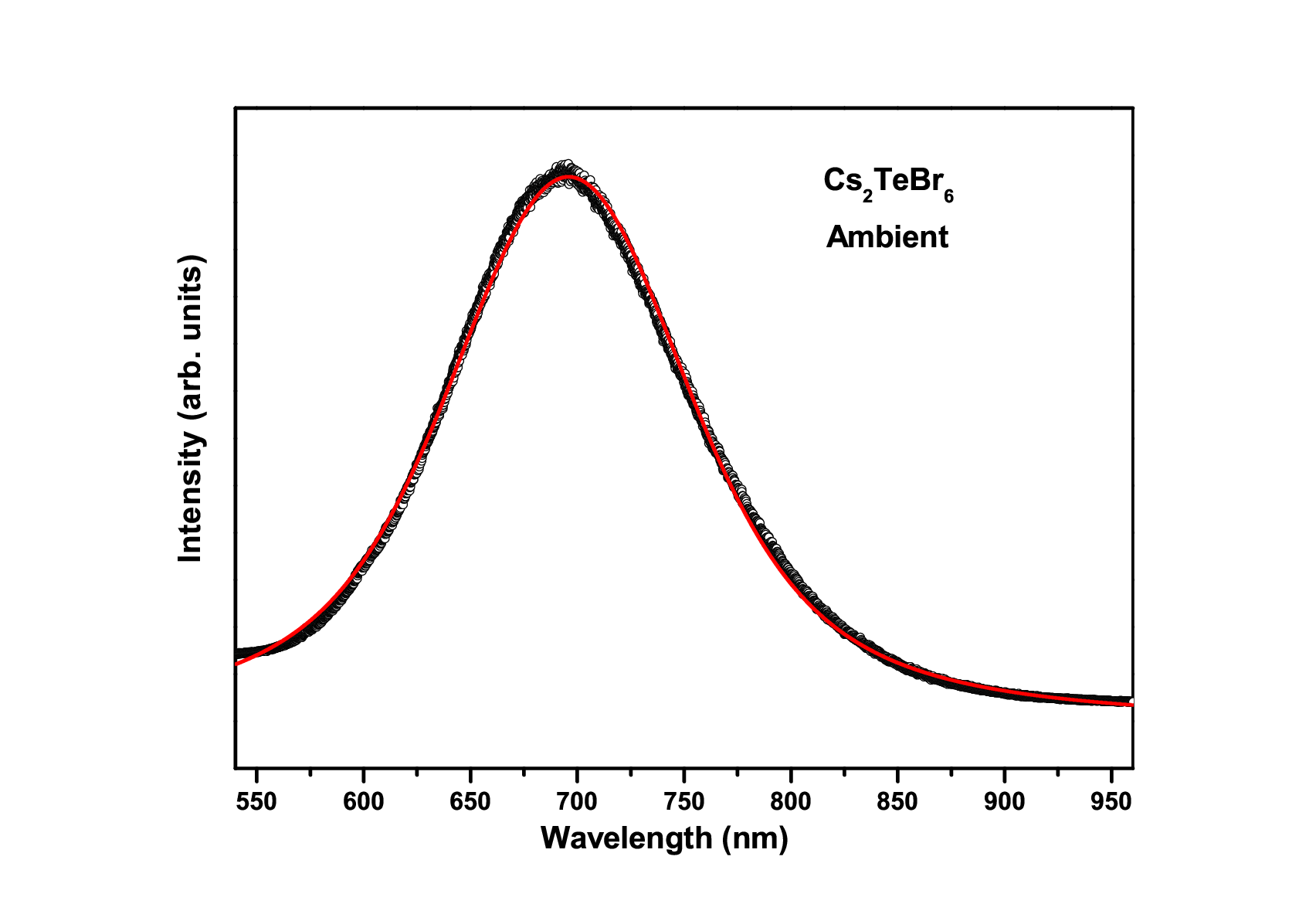}[H]
	\caption*{FIG. S1. Ambient PL spectrum of $Cs_2TeBr_6$. Black circles are experimental data. The experimental data points are fitted to the Voigt function and the fit is shown by the red line.}
\end{figure}

\begin{figure}
	\centering
	\includegraphics[scale = 0.6]{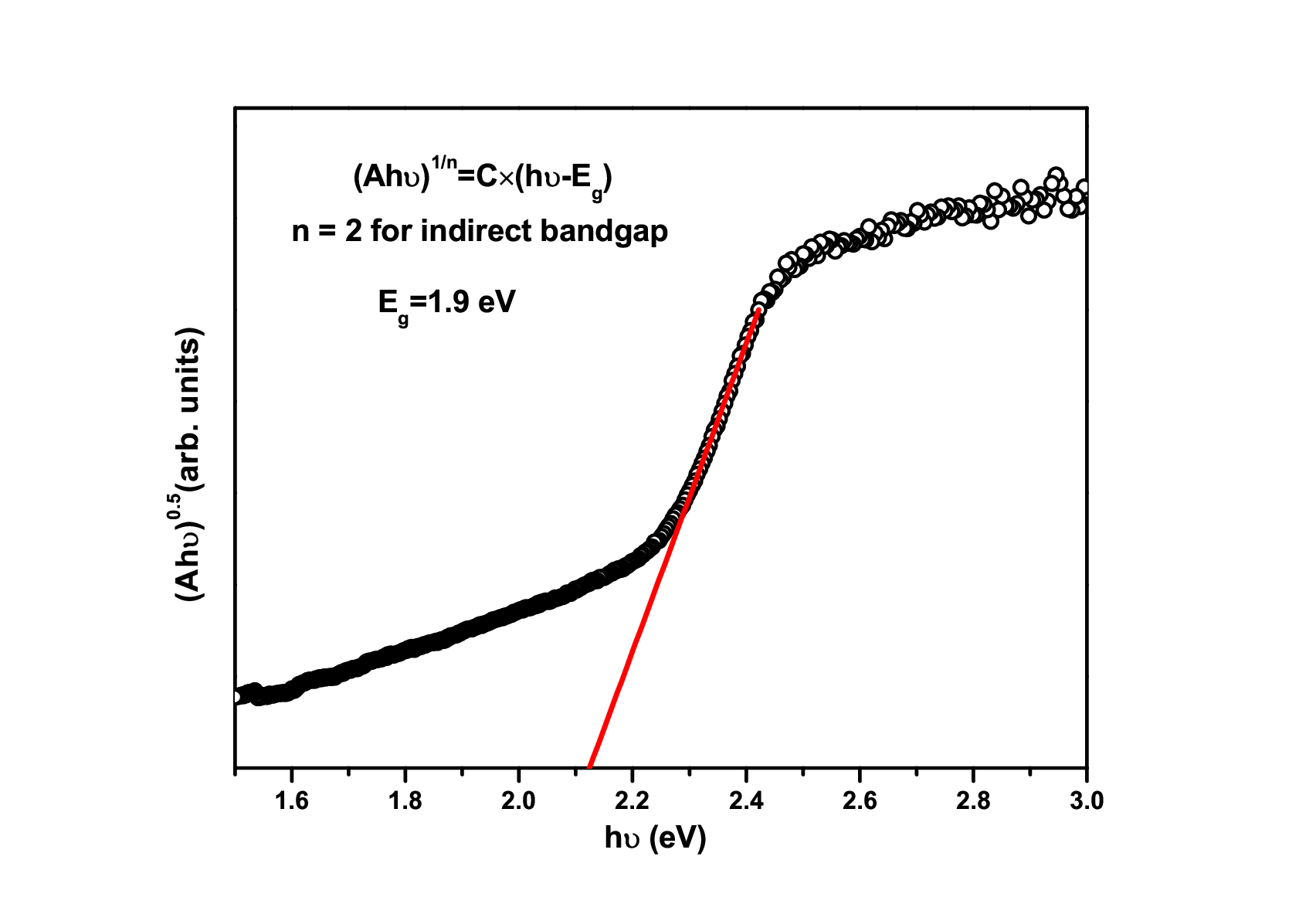}
	\caption*{FIG. S2.  Ambient absorbance spectrum of $Cs_2TeBr_6$.  The bandgap is determined from the absorption edge using a Touc plot for indirect bandgap.  The linear absorption edge is fitted by the equation: $ (A h \nu)^{0.5} =C \times (h \nu -E_g)$;  where $A$ is the absorbance, $h \nu$ is the energy of the photon, $\it C$ is a constant, and $E_g$ is the indirect optical bandgap. The bandgap is measured to be  $E_g$=1.9 eV.}
\end{figure}

\begin{figure}
	\centering
	\includegraphics[scale = 0.6]{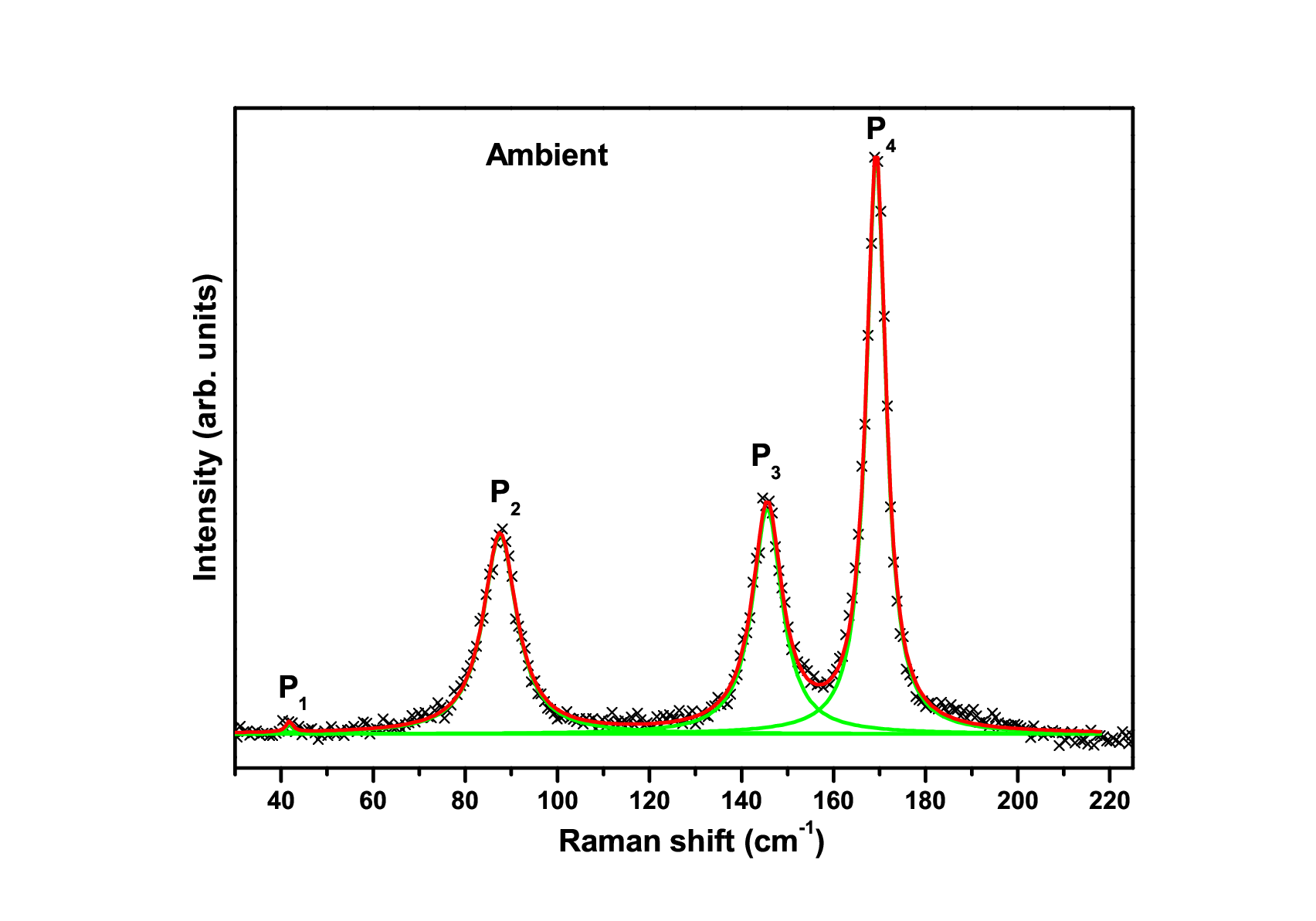}
	\caption*{FIG. S3.Ambient Raman spectrum of $Cs_2TeBr_6$. Black crosses are used to represent the experimental data. The green line shows a fit of experimental data points to Lorentzian functions and the sum of the fit is shown by the red line.}
\end{figure}

\begin{figure}
	\centering
	\includegraphics[scale = 0.6]{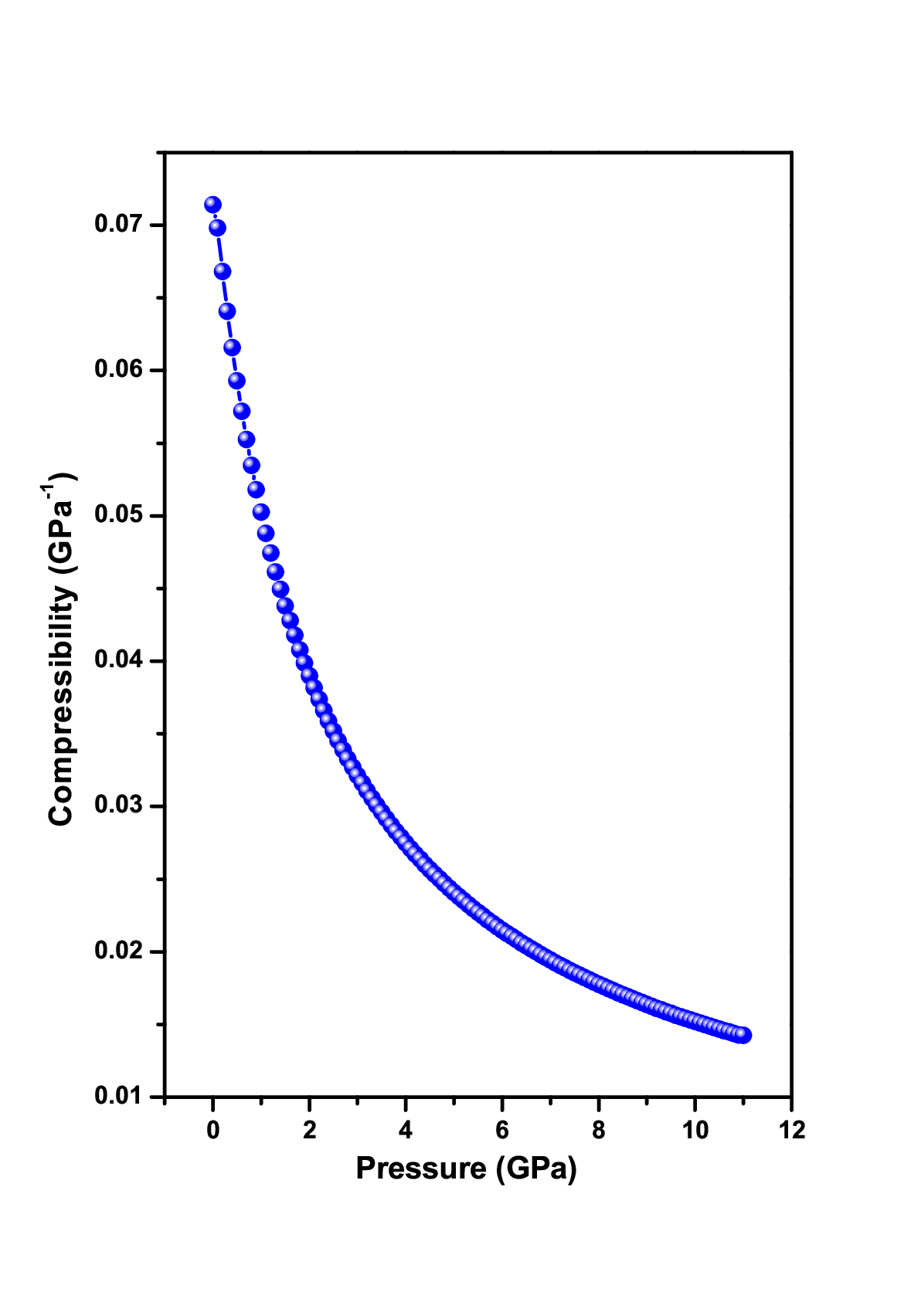}
	\caption*{FIG. S4. A plot of compressibility vs. pressure.}
\end{figure}

    \begin{table}
	\caption*{Table SI. Wyckoff positions, fractional coordinates for the atoms in $Cs_2TeBr_6$ at ambient conditions.} 
	\centering 
	
	\begin{tabular}{c@{\hskip 0.4in} c@{\hskip 0.4in} c@{\hskip 0.4in} c@{\hskip 0.4in} c@{\hskip 0.4in} }
		\hline\hline 
		Atom & Wyckoff position & x & y  & z \\ 
		
		\hline 
		Te& 4a & 0  &  0 &  0 \\
		Cs&  8c & 0.25 & 0.25 & 0.25 \\
		Br& 24e & 0.2490(4)   &   0  &    0 \\ 
		\hline\hline
	\end{tabular}
\end{table}

\end{document}